\documentclass[prb,twocolumn,showpacs,citeautoscript]{revtex4}

\usepackage{graphicx}
\usepackage{amssymb}
\usepackage{amsmath}

\newcommand{\ud}{\,{\mathrm d}}
\newcommand{\zc}{\overline{z}}

\newcommand{\fc}{\overline{f}}

\newcommand{\uiiint}{\int\!\!\!\int\!\!\!\int}
\newcommand{\uRe}{{\mathrm{Re}}\,}

\begin{document}

\title{Magnetization patterns in ferromagnetic nano-elements as functions of complex variable.}

\author{Konstantin L. Metlov}
\affiliation{Donetsk Institute for Physics and Technology NAS, Donetsk, Ukraine 83114}
\email{metlov@fti.dn.ua}
\date{\today}
\begin{abstract}
Assumption of certain hierarchy of soft ferromagnet energy terms, realized in small enough flat nano-elements, allows to obtain explicit expressions for their magnetization distributions. By minimising the energy terms {\em sequentially}, from most to the least important, magnetization distributions are expressed as solutions of Riemann-Hilbert boundary value problem for a function of complex variable. A number of free parameters, corresponding to positions of vortices and anti-vortices, still remain in the expression. These parameters can be found by computing and minimizing the total magnetic energy of the particle with no approximations. Thus, the presented approach is a factory of realistic Ritz functions for analytical (or numerical) micromagnetic calculations. These functions are so versatile, that they may even find applications on their own (e.g. for fitting magnetic microscopy images). Examples are given for multi-vortex magnetization distributions in circular cylinder, and for 2-dimensional domain walls in thin magnetic strips.
\end{abstract}
\maketitle

Complex analysis is a natural language for expressing solutions of many physical problems\cite{Lavrentiev_Shabat} like 2-dimensional electrostatics, planar flows of ideal incompressible fluid, 2-d problems of theory of elasticity, heat flow and, more recently, many-body wave functions\cite{Laughlin1983} of electrons involved in the fractional quantum Hall effect. In all these cases, a certain sets of physical assumptions were identified, permitting to map the complex calculus to a subset of physical problems in a particular domain. The goal of this work is to do the same for nano-magnetics, expressing magnetic structures in flat cylindrical nano-elements via analytic functions of complex variable, employing the conformal mapping to account for element shapes.

\begin{figure}[t]
  \begin{center}
    \includegraphics[width=6cm]{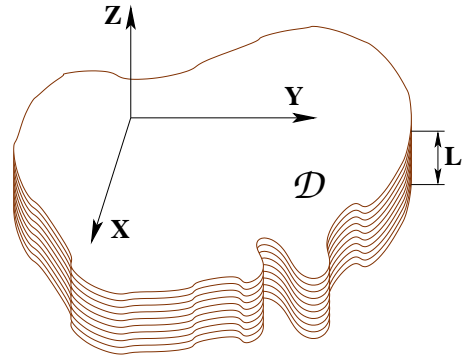}
  \end{center}
  \caption{Cylinder with Cartesian coordinate system axes.}
  \label{fig:cylinder}
\end{figure}
Consider a cylinder with an arbitrary face shape, shown in Fig.~\ref{fig:cylinder} together with Cartesian coordinate system $\vec{r}=\{ X, Y, Z\}$. Its face is denoted by $\cal D$ and its thickness is $L$. Supposing that the cylinder is made of soft ferromagnetic material with saturation magnetization $M_S$ and the exchange stiffness $C$, one can introduce the normalized magnetization vector $\vec{m}(\vec{r})=\{m_X, m_Y, m_Z\}=\vec{M}(\vec{r})/M_S$, $|\vec{m}|=1$ and, classically\cite{Aharoni_book}, express its magnetic energy as
\begin{eqnarray}
  \label{eq:energy_2d}
  \frac{e[\vec{m}]}{\mu_0 M_S^2} = \!\! \uiiint_{{\cal D},L} 
  \left\{
    \frac{L_E^2}{2}\!\!\!\! \sum_{i=X,Y,Z} \!\!\!
    (\vec{\nabla} m_i)^2 -
    \vec{h}_D[\vec{m}]\cdot\vec{m}  
  \right\}\ud^3 \vec{r},
\end{eqnarray}
where $L_E=\sqrt{C/(\mu_0 M_S^2)}$ is the exchange length (of the order of $20nm$ for typical soft magnets), $\vec{\nabla}=\{\partial/\partial X, \partial/\partial Y, \partial/\partial Z\}$, $\vec{h}_D[\vec{m}]$ is demagnetizing field, created by the magnetization distribution $\vec{m}(\vec{r})$. Square brackets denote the {\em functional} dependence (on the whole function, as opposed to its value at a particular point). Both $e$ and $\vec{h}_D$ are functionals of $\vec{m}$.

In the static case with no electric currents the magnetic charge formalism\cite{Aharoni_book} can be used for calculating $\vec{h}_D$ by, first, expressing it as a gradient of a scalar potential function $\vec{h}_D=\vec{\nabla}u(\vec{r})$, which
is, in turn, a solution of the Poisson equation $\vec{\nabla}^2 u = \rho$ with the requirement (due to the finite size of the particle) that both $|\vec{r}| u$ and $|\vec{r}|^2 |\nabla u|$ are finite as $|\vec{r}| \rightarrow \infty$, $\rho=\vec{\nabla}\cdot\vec{m}$ is the density of magnetic charges (on the surface they are proportional to the normal component of $\vec{M}$).

Minimizing the energy functional (\ref{eq:energy_2d}) one would recover both ground and metastable magnetization distributions in the particle. Unfortunately, the general analytical solution of the resulting system of coupled non-linear partial differential integral Euler equations is possible only in a very few cases (most notably for the magnetization states of small ellipsoidal particles\cite{StonerWohlfarth48}). In general case approximations are necessary.

In the following, instead of minimizing all the energy terms simultaneously, the sequential minimization will be adopted. The energy terms are first sorted from the most important to the least important. We also start with the set of all possible functions. Then, in order of decreasing importance of terms, we sieve the current set of functions, keeping only the ones, minimizing (extremalizing, to be precise) the current term. The procedure is repeated with the remaining functions and remaining terms. The question is then how far we can go and how many functions remain at the end ? The final functions, constructed in this way, are, obviously, influenced the most by the first considered (more important) terms.

To reduce the dimensionality of the problem, let us restrict consideration to the particles in the form of planar cylinders, by starting {\em not} from the set of all possible 3-d functions $\vec{m}(\vec{r})$, but from 2-d functions, assuming that the cylinder is thin enough that the magnetization distribution is independent on $Z$, that is $\partial \vec{m}(\vec{r})/\partial Z = 0$. Then it is convenient to introduce the complex coordinate $z=X + \imath Y$, $\imath=\sqrt{-1}$. The set of functions can be sieved further by restricting the length of magnetization vector $|\vec{m}|=1$, defining it via stereographic projection
\begin{eqnarray}
  \label{eq:stereo}
   m_X+\imath m_Y & = & 2 w(z, \zc)/(1+|w(z, \zc)|^2) \\
   \nonumber
   m_Z & = & (1-|w(z, \zc)|^2)/(1+|w(z, \zc)|^2),
\end{eqnarray}
where line over the variable means complex conjugation ($\zc=X - \imath Y$) and the newly introduced complex function of complex variable $w(z, \zc)$ is not necessary holomorphic (that is, not necessarily differentiable) but just an arbitrary relationship between two complex numbers, as noted by its explicit
dependence on $\zc$.

To sort the energy terms, we first note that there are no magnetic monopoles and, thus, the total magnetic charge of the particle (consisting of volume charges, proportional to $\vec{\nabla}\cdot\vec{m}$, and surface charges, proportional to the normal component of $\vec{m}$ at the surface) is zero. As the particle is shrinking, the positive and negative magnetic charges move closer together and so their positive self-energy is more and more compensated by their negative interaction energy. Thus, in small enough particles exchange interaction is more important than magnetostatic. It is more important to minimize the exchange energy even at a cost of having some magnetic charges.

Magnetic charges are also different. There are volume and surface ones. It can be deduced from a very general considerations that for small particles the surface effects dominate. Thus, the volume charges are the least important in our case. Noting that planar cylinders have two surfaces: face and side, let us also distinguish between two corresponding types of surface charges. Of these, the face charges are more important to reduce, because the face is bigger.

Thus, let us assume the following order of sequential minimization: exchange energy, energy of face charges, energy of side charges, and, finally, the energy of volume magnetic charges.

Introducing complex derivatives $\partial/\partial z=(\partial/\partial X - \imath\,
\partial/\partial Y)/2$, $\partial/\partial
\overline{z}=(\partial/\partial X + \imath\, \partial/\partial Y)/2$ the exchange energy density can be represented as
\begin{equation}
  \label{eq:exchange_dens}
 \sum_{i=X,Y,Z} \!\!\!
    (\vec{\nabla} m_i)^2 = \frac{8}{(1+w\overline{w})^2}
  \left(
    \frac{\partial w}{\partial z}
    \frac{\partial \overline{w}}{\partial\overline{z}}+
    \frac{\partial w}{\partial \overline{z}}
    \frac{\partial \overline{w}}{\partial z}
  \right).
\end{equation}
Putting to zero the first variation of integral of this density we get the following Euler equation
\begin{equation}
  \label{eq:energy_euler}
  \frac{\partial}{\partial z}
  \left(\frac{\partial w}{\partial \overline{z}}\right) =
  \frac{2 \overline w}{1+w\overline{w}}
  \frac{\partial w}{\partial z}
  \frac{\partial w}{\partial\overline{z}}.
\end{equation}
This equation is non-linear, it has several families of solutions (their complete set  is still unknown). The first family was found by Belavin and Polyakov\cite{BP75}. These solutions are called ``solitons'' and correspond to $w=f(z)$ being an arbitrary analytic function (that is $\partial w/\partial \zc = 0$). This is obvious in complex notation, in the original work\cite{BP75} derivation was much more involved. There are much more particular solutions of this Euler equation\cite{Hirayama_PLA_1978, Lakshmanan_JPC_1980}, but Belavin and Polyakov solitons are the only ones, having finite energy in infinite ferromagnet\cite{Garber1979}. In restricted thin film geometry other solutions of this Euler equation start to become relevant.

Such other family of solutions was discovered by David Gross\cite{G78}, they are called ``merons'' and expressed as $w(z,\zc)=f(z)/\sqrt{f(z)\fc(\zc)}$, which can be verified by direct substitution. Here $f(z)$ is again an arbitrary analytic function. Merons have $|w|=1$ and, consequently from (\ref{eq:stereo}), $m_z=0$ everywhere. There are no magnetic surface charges on the cylinder faces (and so the magnetostatic energy is at absolute minimum\cite{Aharoni_book}, zero). The exchange energy density of merons, unfortunately, has non-integrable singularities at zeros and poles of $f(z)$, making the exchange energy divergent for any non-trivial $f(z)$. This divergence can be avoided at a cost of some face magnetic charges by joining solitons and merons continuously
\begin{equation}
  \label{eq:sol_SM}
  w(z,\overline{z})=\left\{
    \begin{array}{ll}
      f(z)/e_1 & |f(z)| \leq e_1 \\
      f(z)/\sqrt{f(z) \fc(\zc)} & e_1<|f(z)| \leq e_2\\
      f(z)/e_2 & |f(z)| > e_2
    \end{array}
    \right. ,
\end{equation}
for an arbitrary analytic function $f(z)$ and two arbitrary real constants $0<e_1<e_2<\infty$. Being composed of solutions of Euler equation, this function is locally extremal to the exchange energy functional, while the amount of face magnetic charges can be controlled by selection of the free constants $e_1$ and $e_2$. These constants\cite{M01_solitons} allow to tighten meron arbitrarily close to singularities and zeroes of $f(z)$ trading between the decrease of the magnetostatic energy of face charges and the increase of the exchange energy of the particle.

The next energy term to sieve solutions through is the magnetostatic energy of side charges. Unlike the energy of the face charges, this energy can be completely put to zero (its absolute minimum) without conflict with more important terms we have already minimized. The corresponding requirement\cite{M01_solitons2} for $f(z)$ is {\em to find a function $f(z)$ analytical in the region
  ${\cal D}$ in such a way that $\uRe [ f(\zeta) \overline{n(\zeta)}]
  = 0$ (no magnetization components normal to the side), where $\zeta
  \in {\cal C}=\partial {\cal D}$ is the boundary of ${\cal D}$, and
  $n(\zeta)=n_x (\zeta) +\imath n_y (\zeta)$ is the complex normal to
  ${\cal C}$}. This is an instance of the well known linear Riemann-Hilbert problem\cite{Lavrentiev_Shabat}.

This problem and the resulting magnetization distributions were already considered in Ref.~\onlinecite{M01_solitons2} by reducing the Riemann-Hilbert problem to the problem of Hilbert Privalov on the unit disk, as described in Ref.~\onlinecite{Lavrentiev_Shabat}. This procedure, however, misses some of the physically relevant solutions. To solve the Riemann-Hilbert problem in this work the region ${\cal D}$ is conformally mapped to the upper half plane $t$, such that $\mathrm{Im}\,t > 0$. Then, the functions with non-essential singularities, satisfying condition of no normal component to the boundary of the upper half plane (that is, real on the real axis) can be written, generally, as rational functions with real coefficients
\begin{equation}
\label{eq:ft}
f(t) = \frac{\sum_{i=0}^{m}{g_i t^i}}{\sum_{i=0}^{n}{h_i t^i}},
\end{equation}
where $g_i$ and $h_i$ are arbitrary real numbers. 

In case of arbitrary shape of particle face, let us introduce the conformal mapping $z=M(t)$, transforming the upper half plane $t$ into the particle face $z$. Also note that roots of polynomial with real coefficients are either real or come in complex conjugate pairs. Since the roots of nominator and denominator correspond to zeros and poles of $f(z)$, that is, to soliton ``hats'', covering meron singularities (\ref{eq:sol_SM}), it is  convenient to express these polynomials directly in terms of their roots. Suppose that nominator (denominator) has $m_p$ ($n_p$) pairs of complex roots at points $a_i$, $\overline{a_i}$ ($c_i$, $\overline{c_i}$), such that $\mathrm{Im}\,a_i > 0$ ($\mathrm{Im}\,c_i > 0$); as well as $m_r$ ($n_r$) real roots $b_j$, $\mathrm{Im}\,b_j = 0$ ($d_j$, $\mathrm{Im}\,d_j = 0$). Then $m=2 m_p + m_r$, $n=2n_p+n_r$ and the function $f(z)$ can be written parametrically as
\begin{eqnarray}
\label{eq:fz}
f & = & M'(t) \frac{\prod_{i=0}^{m_p}{(a_i - t) (\overline{a_i} - t)} \prod_{j=0}^{m_r}{(b_j - t)}}
{\prod_{i=0}^{n_p}{(c_i - t) (\overline{c_i} - t)} \prod_{j=0}^{n_r}{(d_j - t)}} \\
\nonumber z & = & M(t).
\end{eqnarray}
This expression together with (\ref{eq:sol_SM}) and (\ref{eq:stereo}) gives the family of Ritz functions for magnetization distributions, depending on a number of parameters $\vec{a}$, $\vec{b}$, $\vec{c}$, $\vec{d}$, $e_1$, $e_2$.

Let us now see some examples. The case of magnetic disk is, probably, the simplest. The most general form of conformal transform of the upper half plane to the unit disk is
\begin{equation}
\label{eq:mtdisk}
M(t) = - \frac{e^{\imath\,\alpha} (h + \imath\,t)}{h - \imath\,t},
\end{equation}
where $h$ is an arbitrary real number and $\alpha\in[0,2\pi)$. In case of $m=2$, $n=0$ this solution coincides with the one, obtained in Ref.~\onlinecite{M01_solitons2}, and further, in a particular case of centered vortex, with an ansatz of Usov and Peschany\cite{UP93}. For bigger $m$ and $n$ a particular multi-vortex magnetization distribution in a cylinder is shown in 
\begin{figure}[t]
  \begin{center}
    \includegraphics[width=7cm]{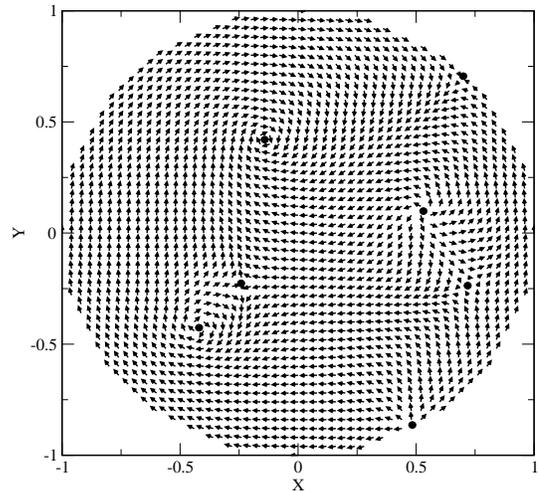}
  \end{center}
  \caption{A multi-vortex magnetization distribution in circular cylinder, comparable to the numerical simulations of Ref.~\onlinecite{Kim2004}. It is (\ref{eq:stereo}), (\ref{eq:sol_SM}), (\ref{eq:fz}), (\ref{eq:mtdisk}) with $h=4$, $\alpha=\pi/4$, $e_1=0.01$, $e_2=400$, $(\vec{a}, \vec{b}, \vec{c}, \vec{d})=(\{\imath,-4+4\imath, 6+7\imath\},\{3\},\{2\imath,6+2\imath\},\{\})$. There are three vortices, two anti-vortices and two side-bound skyrmions, their centers, where the magnetization vector is vertical, are marked with dots.}
  \label{fig:disk}
\end{figure}
Fig.~\ref{fig:disk} with parameters, chosen for plausibility.

The other interesting case is that of a 2-dimensional domain wall in thin strip. This problem had been treated numerically\cite{Donahue_VortexWall1997}, but has resurfaced recently in connection with the idea of racetrack magnetic memory\cite{ParkinRacetrack2008}. Noting that the conformal map from the upper half plane to the infinite strip $0<\mathrm{Im}\,z<1$ is
\begin{equation}
\label{eq:mtstrip}
M(t) = - \frac{\log t}{\pi},
\end{equation}
where additional parameters, related to a certain freedom to transform the upper half plane to itself, were omitted for clarity. Some of possible domain wall configurations are shown in 
\begin{figure}[t]
  \begin{center}
    \includegraphics[width=7.8cm]{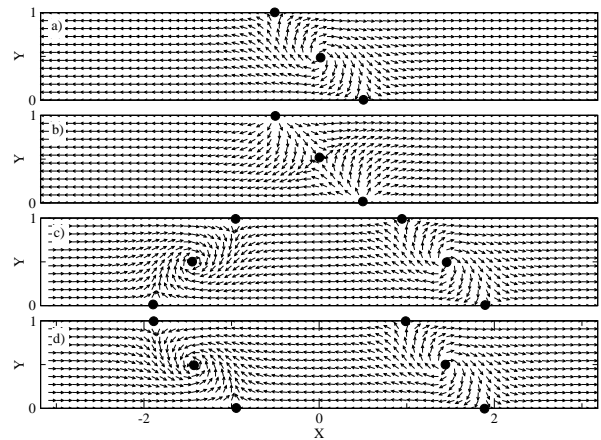}
  \end{center}
  \caption{Some domain wall configurations in thin strip: a) vortex wall (comparable to the numerical simulation of Ref.~\onlinecite{Donahue_VortexWall1997}); b) anti-vortex wall; c) and d) two different arrangements of a pair of two vortex walls. As before, dots mark the singularities in the meron. It is (\ref{eq:stereo}), (\ref{eq:sol_SM}), (\ref{eq:fz}), (\ref{eq:mtstrip}) with $e_1=0.01$, $e_2=400$, and $(\vec{a}, \vec{b}, \vec{c}, \vec{d})$: a) $(\{I\},\{0\},\{\},\{5,-1/5\})$; b)  $(\{\},\{0, 5, -1/5\},\{I\},\{\})$; c)  $(\{90 \imath, \imath/90\},\{0\},\{\},\{360, 1/360, -20, -1/20 \})$; d)  $(\{90 \imath, \imath/90\},\{0\},\{\},\{360, -1/360, -20, 1/20 \})$.}
  \label{fig:walls}
\end{figure}
Fig.~\ref{fig:walls}. It is trivial to extend these expressions to model the whole racetrack, containing many interacting domain walls of different types.

Concluding the step by step minimization procedure, we have sieved the set of all possible magnetization distributions to obtain a much smaller set, parametrized by a finite number of scalar parameters. The values of these remaining parameters, corresponding to ground and metastable states of magnetization, can be found by minimizing the total energy (\ref{eq:energy_2d}). This problem is much simpler, compared to full solution of corresponding Euler equations, and is tractable analytically in a number of interesting cases, e.g. like that of displaced magnetic vortex\cite{MG02_JEMS} or of ``C''-type low-symmetry magnetic states in circular cylinders\cite{ML08}.

It is also possible to generalize these Ritz functions by relaxing the strict requirement of no normal magnetization components introducing an additional parameter, corresponding to rescaling the particle. This works very well for treatment of quasi-uniform states in circular cylinders\cite{MG04}, reproducing  quite intricate measurements\cite{M06}. This generalization can be applied to strips (by rescaling their width) and should produce a very precise set of Ritz functions.

There is no ``real'' unchangeable by evolution topological charge in finite particles, since the energy barrier, separating the states with different polynomial degrees of nominator or doniminator in (\ref{eq:fz}), is finite (unlike the case of infinite ferromagnet, considered by Belavin and Polyakov). Nevertheless, since the exchange interaction is ``the most important'' in the particles of a few exchange lengths in size, the difference $m-n$ has a pronounced effect on the magnetic states and still remains useful for the purpose of classification even though there is no direct proportionality between the total energy and $|m-n|$ as in Belavin-Polyakov case.

While the exchange energy of magnetization distributions (\ref{eq:fz}), consisting both of soliton and meron parts, can be calculated analytically in  general form by Greene and residue theorems, magnetostatic energy in (\ref{eq:energy_2d}) currently has to be calculated on a case by case basis. Finding generic expression for magnetostatic energy of magnetization distributions (\ref{eq:fz}) is still an open problem. Absence of such an expression and long-range character of dipolar interaction (resulting in substantial dependence of its energy on particle shape) makes it difficult to estimate the relative importance of energy terms {\it a priori} for certain particle size. As with other approaches, involving trial functions, the rigorous statement is that the energy of these approximate solutions if always higher than that of exact ones. This allows to check the validity of obtained results {\it a posteriori}. As a rule of thumb, for particles with the dimensions of a few $L_E$ the magnetic states, their energies and various derivatives are usually in a good agreement to experiment. For larger particles results may vary.

Functions (\ref{eq:fz}) are so versatile that they might be useful on their own, beyond the Ritz method.  For example, to enhance the resolution of magnetic force microscopy images of small particles by fitting the positions of topological singularities, comparing calculated and measured forces on the tip.

Magnetization dynamics can be considered separately\cite{GAG2010} on top of these distributions.

The author would like to thank Andrei Bogatyrev from the Institute of Numerical Methods of Russian Academy of Sciences for his help with solution of Riemann-Hilbert problem, which have eventually led to the expression (\ref{eq:ft}).


\begin{thebibliography}{20}
\expandafter\ifx\csname natexlab\endcsname\relax\def\natexlab#1{#1}\fi
\expandafter\ifx\csname bibnamefont\endcsname\relax
  \def\bibnamefont#1{#1}\fi
\expandafter\ifx\csname bibfnamefont\endcsname\relax
  \def\bibfnamefont#1{#1}\fi
\expandafter\ifx\csname citenamefont\endcsname\relax
  \def\citenamefont#1{#1}\fi
\expandafter\ifx\csname url\endcsname\relax
  \def\url#1{\texttt{#1}}\fi
\expandafter\ifx\csname urlprefix\endcsname\relax\def\urlprefix{URL }\fi
\providecommand{\bibinfo}[2]{#2}
\providecommand{\eprint}[2][]{\url{#2}}

\bibitem[{\citenamefont{Lavrentiev and Shabat}(1965)}]{Lavrentiev_Shabat}
\bibinfo{author}{\bibfnamefont{M.~A.} \bibnamefont{Lavrentiev}}
  \bibnamefont{and} \bibinfo{author}{\bibfnamefont{B.~V.}
  \bibnamefont{Shabat}}, \emph{\bibinfo{title}{Methods of the theory of
  functions of complex variable}} (\bibinfo{publisher}{Nauka},
  \bibinfo{address}{Moskva}, \bibinfo{year}{1965}).

\bibitem[{\citenamefont{Laughlin}(1983)}]{Laughlin1983}
\bibinfo{author}{\bibfnamefont{R.~B.} \bibnamefont{Laughlin}},
  \bibinfo{journal}{Phys. Rev. Lett.} \textbf{\bibinfo{volume}{50}},
  \bibinfo{pages}{1395} (\bibinfo{year}{1983}).

\bibitem[{\citenamefont{Aharoni}(1996)}]{Aharoni_book}
\bibinfo{author}{\bibfnamefont{A.}~\bibnamefont{Aharoni}},
  \emph{\bibinfo{title}{Introduction to the theory of ferromagnetism}}
  (\bibinfo{publisher}{Oxford University Press}, \bibinfo{address}{Oxford},
  \bibinfo{year}{1996}).

\bibitem[{\citenamefont{Stoner and Wohlfarth}(1948)}]{StonerWohlfarth48}
\bibinfo{author}{\bibfnamefont{E.~C.} \bibnamefont{Stoner}} \bibnamefont{and}
  \bibinfo{author}{\bibfnamefont{E.~P.} \bibnamefont{Wohlfarth}},
  \bibinfo{journal}{Phil. Trans. R. Soc. Lond. A}
  \textbf{\bibinfo{volume}{240}}, \bibinfo{pages}{599} (\bibinfo{year}{1948}).

\bibitem[{\citenamefont{Belavin and Polyakov}(1975)}]{BP75}
\bibinfo{author}{\bibfnamefont{A.~A.} \bibnamefont{Belavin}} \bibnamefont{and}
  \bibinfo{author}{\bibfnamefont{A.~M.} \bibnamefont{Polyakov}},
  \bibinfo{journal}{ZETP lett.} \textbf{\bibinfo{volume}{22}},
  \bibinfo{pages}{503} (\bibinfo{year}{1975}), \bibinfo{note}{(in Russian)}.

\bibitem[{\citenamefont{Hirayama et~al.}(1978)\citenamefont{Hirayama, Tze,
  Ishida, and Kawabe}}]{Hirayama_PLA_1978}
\bibinfo{author}{\bibfnamefont{M.}~\bibnamefont{Hirayama}},
  \bibinfo{author}{\bibfnamefont{H.~C.} \bibnamefont{Tze}},
  \bibinfo{author}{\bibfnamefont{J.}~\bibnamefont{Ishida}}, \bibnamefont{and}
  \bibinfo{author}{\bibfnamefont{T.}~\bibnamefont{Kawabe}},
  \bibinfo{journal}{Phys. Lett. A} \textbf{\bibinfo{volume}{66}},
  \bibinfo{pages}{352 } (\bibinfo{year}{1978}).

\bibitem[{\citenamefont{Lakshmanan et~al.}(1980)\citenamefont{Lakshmanan,
  Daniel, and Kaliappan}}]{Lakshmanan_JPC_1980}
\bibinfo{author}{\bibfnamefont{M.}~\bibnamefont{Lakshmanan}},
  \bibinfo{author}{\bibfnamefont{M.}~\bibnamefont{Daniel}}, \bibnamefont{and}
  \bibinfo{author}{\bibfnamefont{P.}~\bibnamefont{Kaliappan}},
  \bibinfo{journal}{J. Phys. C} \textbf{\bibinfo{volume}{13}},
  \bibinfo{pages}{4743 } (\bibinfo{year}{1980}).

\bibitem[{\citenamefont{Garber et~al.}(1979)\citenamefont{Garber, Ruijsenaars,
  Seiler, and {an Burns}}}]{Garber1979}
\bibinfo{author}{\bibfnamefont{W.-D.} \bibnamefont{Garber}},
  \bibinfo{author}{\bibfnamefont{S.~N.~M.} \bibnamefont{Ruijsenaars}},
  \bibinfo{author}{\bibfnamefont{E.}~\bibnamefont{Seiler}}, \bibnamefont{and}
  \bibinfo{author}{\bibfnamefont{D.}~\bibnamefont{{an Burns}}},
  \bibinfo{journal}{Annals of Physics} \textbf{\bibinfo{volume}{119}},
  \bibinfo{pages}{305 } (\bibinfo{year}{1979}).

\bibitem[{\citenamefont{Gross}(1978)}]{G78}
\bibinfo{author}{\bibfnamefont{D.~J.} \bibnamefont{Gross}},
  \bibinfo{journal}{Nucl. Phys. B} \textbf{\bibinfo{volume}{132}},
  \bibinfo{pages}{439} (\bibinfo{year}{1978}).

\bibitem[{\citenamefont{Metlov}(2000)}]{M01_solitons}
\bibinfo{author}{\bibfnamefont{K.~L.} \bibnamefont{Metlov}}
  (\bibinfo{year}{2000}), \bibinfo{note}{{\tt arXiv:cond-mat/0012146}}.

\bibitem[{\citenamefont{Metlov}(2001)}]{M01_solitons2}
\bibinfo{author}{\bibfnamefont{K.~L.} \bibnamefont{Metlov}}
  (\bibinfo{year}{2001}), \bibinfo{note}{{\tt arXiv:cond-mat/0102311}}.

\bibitem[{\citenamefont{Usov and Peschany}(1993)}]{UP93}
\bibinfo{author}{\bibfnamefont{N.~A.} \bibnamefont{Usov}} \bibnamefont{and}
  \bibinfo{author}{\bibfnamefont{S.~E.} \bibnamefont{Peschany}},
  \bibinfo{journal}{J. Magn. Magn. Mater.} \textbf{\bibinfo{volume}{118}},
  \bibinfo{pages}{L290} (\bibinfo{year}{1993}).

\bibitem[{\citenamefont{Lee et~al.}(2004)\citenamefont{Lee, Kang, Yu, and
  Kim}}]{Kim2004}
\bibinfo{author}{\bibfnamefont{K.-S.} \bibnamefont{Lee}},
  \bibinfo{author}{\bibfnamefont{B.-W.} \bibnamefont{Kang}},
  \bibinfo{author}{\bibfnamefont{Y.-S.} \bibnamefont{Yu}}, \bibnamefont{and}
  \bibinfo{author}{\bibfnamefont{S.-K.} \bibnamefont{Kim}},
  \bibinfo{journal}{Appl. Phys. Lett.} \textbf{\bibinfo{volume}{85}},
  \bibinfo{pages}{1568} (\bibinfo{year}{2004}).

\bibitem[{\citenamefont{McMichael and Donahue}(1997)}]{Donahue_VortexWall1997}
\bibinfo{author}{\bibfnamefont{R.}~\bibnamefont{McMichael}} \bibnamefont{and}
  \bibinfo{author}{\bibfnamefont{M.}~\bibnamefont{Donahue}},
  \bibinfo{journal}{IEEE Trans. Magn.} \textbf{\bibinfo{volume}{33}},
  \bibinfo{pages}{4167} (\bibinfo{year}{1997}).

\bibitem[{\citenamefont{Parkin et~al.}(2008)\citenamefont{Parkin, Hayashi, and
  Thomas}}]{ParkinRacetrack2008}
\bibinfo{author}{\bibfnamefont{S.~S.~P.} \bibnamefont{Parkin}},
  \bibinfo{author}{\bibfnamefont{M.}~\bibnamefont{Hayashi}}, \bibnamefont{and}
  \bibinfo{author}{\bibfnamefont{L.}~\bibnamefont{Thomas}},
  \bibinfo{journal}{Science} \textbf{\bibinfo{volume}{320}},
  \bibinfo{pages}{190} (\bibinfo{year}{2008}).

\bibitem[{\citenamefont{Metlov and Guslienko}(2002)}]{MG02_JEMS}
\bibinfo{author}{\bibfnamefont{K.~L.} \bibnamefont{Metlov}} \bibnamefont{and}
  \bibinfo{author}{\bibfnamefont{K.~Y.} \bibnamefont{Guslienko}},
  \bibinfo{journal}{J. Magn. Magn. Mater.} \textbf{\bibinfo{volume}{242--245}},
  \bibinfo{pages}{1015} (\bibinfo{year}{2002}).

\bibitem[{\citenamefont{Metlov and Lee}(2008)}]{ML08}
\bibinfo{author}{\bibfnamefont{K.~L.} \bibnamefont{Metlov}} \bibnamefont{and}
  \bibinfo{author}{\bibfnamefont{Y.~P.} \bibnamefont{Lee}},
  \bibinfo{journal}{Appl. Phys. Lett.} \textbf{\bibinfo{volume}{92}},
  \bibinfo{pages}{112506} (\bibinfo{year}{2008}).

\bibitem[{\citenamefont{Metlov and Guslienko}(2004)}]{MG04}
\bibinfo{author}{\bibfnamefont{K.~L.} \bibnamefont{Metlov}} \bibnamefont{and}
  \bibinfo{author}{\bibfnamefont{K.~Y.} \bibnamefont{Guslienko}},
  \bibinfo{journal}{Phys. Rev. B} \textbf{\bibinfo{volume}{70}},
  \bibinfo{pages}{052406} (\bibinfo{year}{2004}).

\bibitem[{\citenamefont{Metlov}(2006)}]{M06}
\bibinfo{author}{\bibfnamefont{K.~L.} \bibnamefont{Metlov}},
  \bibinfo{journal}{Phys. Rev. Lett.} \textbf{\bibinfo{volume}{97}},
  \bibinfo{pages}{127205} (\bibinfo{year}{2006}).

\bibitem[{\citenamefont{Guslienko et~al.}(2010)\citenamefont{Guslienko, Aranda,
  and Gonzalez}}]{GAG2010}
\bibinfo{author}{\bibfnamefont{K.~Y.} \bibnamefont{Guslienko}},
  \bibinfo{author}{\bibfnamefont{G.~R.} \bibnamefont{Aranda}},
  \bibnamefont{and} \bibinfo{author}{\bibfnamefont{J.~M.}
  \bibnamefont{Gonzalez}}, \bibinfo{journal}{Phys. Rev. B}
  \textbf{\bibinfo{volume}{81}}, \bibinfo{pages}{014414}
  (\bibinfo{year}{2010}).

\end{thebibliography}

\end{document}